\documentstyle[twoside,fleqn,espcrc2,epsf]{article}

\newcommand{\AmS}{{\protect\the\textfont2
  A\kern-.1667em\lower.5ex\hbox{M}\kern-.125emS}}
\newcommand{\eq}[1]{(\ref{#1})}
\newcommand{\beq}{\begin{equation}}
\newcommand{\eeq}{\end{equation}}
\newcommand{\beqn}{\begin{eqnarray}}
\newcommand{\eeqn}{\end{eqnarray}}

\newcommand{\cC}{{\cal{C}}}
\newcommand{\cZ}{{\cal{Z}}}
\newcommand{\cX}{{\cal{X}}}

\newcommand{\LL}{{I\!\! L}}

\def\NP{ Nucl.~Phys.}

\def\PRL{ Phys.~Rev.~Lett.}
\def\PRp{ Phys.~Rep.}
\def\PR{ Phys.~Rev.}

\title{\vspace{-2.cm}
\begin{flushright}
KANAZAWA 96-19\\
August 1996
\end{flushright}
\vspace{.2cm}
Recent Results on the Abelian Projection of Lattice
Gluodynamics}

\author{M.I.~Polikarpov{\address{Department of Physics, Kanazawa
University, Kanazawa 920-11, Japan\\
\mbox{and} ITEP, B.Cheremushkinskaya 25, Moscow, 117259 Russia}%
}}

\begin{document}

\begin{abstract}
The abelian projection of lattice gluodynamics is reviewed. The main
topics are: abelian and monopole dominance, monopole condensate as the
disorder parameter, effective abelian Lagrangian, monopoles in the
instanton field, Aharonov -- Bohm effect on the lattice.
\end{abstract}

\maketitle

\section{INTRODUCTION}

We do not have much intuition in nonabelian theories,
whereas our understanding of the
Maxwell equations and abelian theories
is far better. For this reason
people are trying to explain the
confinement phenomenon in terms of the abelian projection of
nonabelian theory \cite{tHo81}. Different abelian projections lead to
different abelian theories, and now it is clear that there exist at
least one abelian projection, called the maximal abelian projection (MA)
\cite{KrScWi87}, in which the resulting abelian theory is close to the
dual Abelian Higgs model, the abelian monopoles are condensed and the
linear $q-\bar{q}$ potential can be explained \cite{MatH76} at the
level of the classical equations of motion (formation of the dual
Abrikosov vortex). The dependence of the abelian theory on the type of
the projection is a weak point of this approach, and the popular idea
is that all abelian projections can lead to the same physics, if we
consider a certain generalized version of the abelian projection,
e.g. extended monopoles.

Most of the numerical results are obtained in the MA projection which,
for the $SU(2)$ gauge theory, corresponds to the minimization of the
functional:

\beq
R = \int \{ [A^1_\mu(x)]^2 + [A^2_\mu(x)]^2 \} \, d^4 x \, . \label{R}
\eeq
The corresponding differential equation is:

\beq
[\partial_\mu \pm i e A^3_\mu(x)] A^\pm_\mu(x) = 0 \, . \label{dR}
\eeq

In Section 2 the abelian and monopole dominance is discussed with
regard to various
abelian projections. In Section 3 we discuss
the effective monopole Lagrangian
and the phenomenon of the monopole condensation. The
relation of monopoles and instantons is reviewed in Section 4. In
Section 5 it is shown that an analogue of the Aharonov -- Bohm effect
can be found in lattice gauge theories.

\section{ABELIAN AND MONOPOLE DOMINANCE}
\subsection{Maximal Abelian Projection}
The notion of the ``abelian dominance'' introduced in
\cite{SuYo90} means that the expectation value of the physical
quantity $<\cX>$ in nonabelian theory coincides with the corresponding
expectation value in the abelian theory obtained by the abelian
projection. The monopole dominance means that the same quantity can
be calculated in terms of the monopole currents extracted from the
abelian fields. If we have $N$ configurations of the nonabelian
fields on the lattice, the abelian dominance means that:

\beqn
\frac1N\sum_{conf}\cX(\hat U_{nonabelian})= \nonumber \\
\frac1N\sum_{conf}\cX '(U_{abelian})=
\frac1N\sum_{conf}\cX ''(j) \, . \label{abdom}
\eeqn
Here each sum is taken over all configurations;
$U_{abelian} = e^{i\theta_l}$ is the abelian part of the nonabelian
field $\hat U_{nonabelian}$, $j$ is the monopole current extracted
from $U_{abelian}$.  It is clear that $\frac1N\sum_{conf}\cX(\hat
U_{nonabelian})$ is a gauge invariant quantity, while the abelian
and the monopole contributions depend on the type of the abelian
projection. In numerical calculations the equalities \eq{abdom} can
only be satisfied approximately.

Among the well-studied problems  is that  of  the abelian and the monopole
dominance for
the string tension \cite{SuYo90,HiKiKi91,StNeWe94,BaBoMu96}. In this case,
$\cX(\hat U_{nonabelian}) = \sigma_{SU(2)}$, $\cX(U_{abelian}) =
\sigma_{U(1)}$ and the string tension $\sigma_{SU(2)}$
($\sigma_{U(1)}$) is calculated by means of the nonabelian (abelian)
Wilson loops, $Tr \prod_{l\in C} \hat U_l$ ( $\prod_{l\in C}
e^{i\theta_l}$). An accurate numerical study of the MA projection of
$SU(2)$ gluodynamics on $32^4$ lattice at $\beta = 2.5115$ is
performed in ref.\cite{BaBoMu96}. The abelian and the nonabelian
potentials are shown in Fig.~1 (taken from \cite{BaBoMu96}). The
contribution of the photon and the monopole parts to the abelian
potential is shown in Fig.~2 \cite{BaBoMu96}.

\begin{figure}[htb]
\vspace{-.5cm} 
\centerline{\epsfxsize=0.45\textwidth\epsfbox{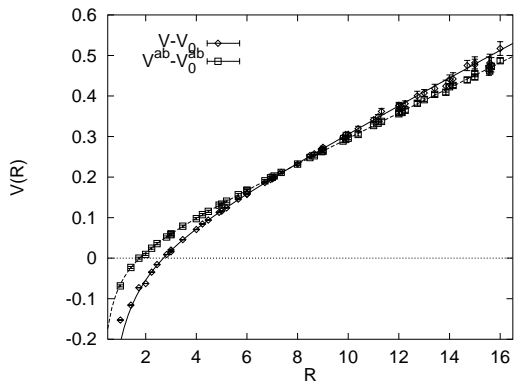}} 
\vspace{-1.0cm} 
\caption{Abelian and nonabelian potentials}
\vspace{-0.6cm} 
\end{figure}

\begin{figure}[htb]
\vspace{-0.6cm}
\centerline{\epsfxsize=0.45\textwidth\epsfbox{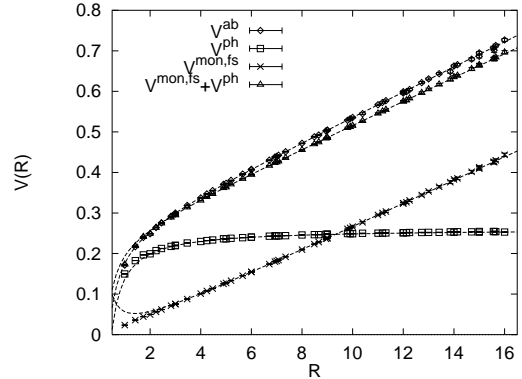}}
\vspace{-1.0cm}
\caption{The abelian potential (diamonds) in comparison
with the photon contribution (squares), the monopole contribution
(crosses) and the sum of these two parts (triangles).}
\vspace{-0.6cm}
\end{figure}

The differences in the slopes of the linear part of the potentials in
Fig.~1 and Fig.~2 yield the following relations: $\sigma_{U(1)}
\approx 92\% \, \sigma_{SU(2)}$, $\sigma_j \approx 95\% \,
\sigma_{U(1)}$, where $\sigma_j$ is the monopole current contribution
to the string tension. It is important to study a widely discussed
idea that in the continuum limit ($\beta \to \infty$) the abelian and
the monopole dominance is exact \eq{abdom}: $\sigma_{SU(2)} =
\sigma_{U(1)} = \sigma_j$.

There are many examples of the abelian and the monopole dominance in
the MA projection.  The monopole dominance for the string tension has
been found in the $SU(2)$ positive plaquette model in which $Z_2$
monopoles are suppressed \cite{StNe96}, it has also been found for the
$SU(2)$ string tension at finite temperature \cite{StWeNe96} and for
the string tension in the $SU(3)$ gluodynamics \cite{Yee94}. The
abelian and the monopole dominance for the $SU(2)$ gluodynamics has
been found in \cite{Kit96,Eji96} for the Polyakov line and for the
critical exponents for the Polyakov line, for the value of the quark
condensate, for the topological susceptibility and also for the hadron
masses in the quenched $SU(3)$ QCD with Wilson fermions \cite{Suzm96}.

\subsection{Two Open Problems}

Consider the adjoint sources in the $SU(2)$ gluodynamics. After the
abelian projection, the corresponding Wilson loop can be represented
as a sum of the charge two, $W_2$, and the charge zero, $W_0$,
contributions. This fact give rise to a paradox discussed in
refs. \cite{DeFaGr96,Gre96,Pou96,Poul96}: for the zero $U(1)$ charge
there is no confinement (therefore there is no area low for the sum
$W_2 + W_0$), but for the $SU(2)$ gluodynamics at the intermediate
distances there exists string tension for the adjoint sources
(``Casimir scaling'' \cite{DeFaGr96}). A partial solution of this
problem has been found in refs. \cite{Pou96,Poul96}: it is
possible to reproduce the $SU(2)$ string tension for the adjoint
sources using only abelian variables (see Fig.11 of ref.\cite{Pou96}
and Fig.3 of ref.\cite{Poul96}).

Another widely discussed question is that, possibly, the property of
the abelian dominance can be proved analytically, and is, therefore,
trivial. The idea is as follows. Consider an irreducible correlator
$G(x) = {<A(x)A(0)>}$. By the spectral theorem $G(x) \approx C
e^{- m |x|}$ at large $|x|$; the constant C depends on the choice of
$A$, but the mass $m$ is the same for all $A$ with the same quantum
numbers. Similar arguments based on the transfer matrix approach
show that the string tension should be the same for the $SU(2)$
sources in the full $SU(2)$ theory and for the $U(1)$ sources after
the abelian projection. But for the MA projection the resulting $U(1)$
theory is nonlocal in space and in time and the use of the transfer
matrix approach and/or spectral theorems is questionable.

\subsection{Various Abelian Projections}

Different abelian projections lead to different $U(1)$ gauge
theories. The ``extreme'' example is the ``minimal abelian
projection''
\cite{ChVePo95}, in which the properties of the monopole currents differ
much from those in the MA projection. The projection in which the field
strength $\hat F_{12}$ is diagonalized also yields the results which
are different from those in the MA projection
\cite{IvPoPo90,DiGi95,HaTe96}. The projection, corresponding to
the diagonalization of the Polyakov line is closer to the MA
projection, but still the results are not the same
\cite{DiGi95,Hay96,HaTe96}. For instance, the Abrikosov vortices are
suppressed in this projection compared to the MA projection \cite{Hay96}.
There are two new examples of the abelian projection (mAA and mAMD)
\cite{Sul96} with the results very close to the MA projection:
the abelian and the monopole dominance is found in these projections
(see Fig. 5 of ref.\cite{Sul96}). Due to the presence of the Gribov
copies it is difficult to fix numerically the MA projection, just as
the projections suggested in ref.\cite{Sul96}. A smooth abelian
projection which is free from the Gribov copies is suggested in
ref.\cite{vdS96}, a numerical study shows that the properties of the
monopole currents in this projection are close to the properties of
the monopole currents in the MA projection; the abelian dominance has
not yet been studied.

In ref.\cite{Gre96} the ``Maximal $Z_2$ gauge'' is suggested in which
the $SU(2)$ fields are projected on the $Z_2$ fields. It is
instructive that the approximate abelian dominance (center dominance)
for the string tension exists in this projection. This fact allows one
to discuss
\cite{Gre96} the ``spaghetti vacuum model'' as the supplementary model
to the model of the dual superconducting vacuum. Another observation
from ref.\cite{Gre96} should be mentioned: if we perform the MA
projection only for the gauge fields $U_{x,x+\hat \mu}$ with $\mu =
1,2$ and do not fix the MA projection for $\mu = 3,4$, then the
abelian dominance takes place for the string tension constructed from Wilson 
loops in $1-2$ plane.

\section{MONOPOLE CONDENSATE AND EFFECTIVE MONOPOLE ACTION}

\subsection{Disorder Parameter for the Deconfinement Phase Transition}

If the vacuum of the $SU(2)$ gluodynamics in the abelian projection is
similar to the dual superconductor, then the value of the monopole
condensate should depend on the temperature as a disorder parameter:
at low temperatures it should be nonzero, and it should vanish above the
deconfinement phase transition. A numerical study of the monopole condensate
has been recently performed by three teams
\cite{DiGi95,DiGir96,ChPoVeo95,Nak96}.

The logarithmic derivative of the
monopole creation operator with respect to $\beta$ ($\rho = \partial
\varphi/ \partial\beta$)  for $SU(2)$ lattice gluodynamics is studied in
refs.\cite{DiGi95,DiGir96}, where it is shown to have a peak just at
the point of the phase transition. A similar operator exhibits the same
behavior in the lattice compact electrodynamics, in the $SU(3)$ lattice
gauge theory, and in the $3D$ $XY$ model \cite{DiGir96,CeGiPa96}.

Another form of the monopole creation operator is studied in
ref.\cite{ChPoVeo95}. This form is similar to that suggested by Fr\"ohlich
and Marchetti \cite{FrMa86} for the compact electrodynamics. For the
$SU(2)$ lattice gluodynamics in the MA projection, it is convenient to
study the probability distribution of the value of the monopole
creation operator (similar calculations were performed for the
compact electrodynamics in ref. \cite{PoPoWi91}). It occurs that at
low temperatures, below the deconfinement phase transition the maximum
of the distribution is shifted from zero, which means that the
effective constraint potential is of the Higgs type. Above the phase
transition the minimum of the potential, $\varphi_C$, (the maximum of
the monopole field distribution) is at the zero value of the monopole
field. The dependence of the quantity $\varphi_C$ (which is
proportional to the value of the monopole condensate) on $\beta$ is
shown in Fig.~3. It is obvious that $\varphi_C$ behaves as the
disorder parameter. To get this result the calculations are performed
on the lattices of the size $4\times 8^3,\, 4\times 10^3,\,4\times
12^3,\,4\times 14^3, \,4\times 16^3$ and the data for $\varphi_C$ are
extrapolated to the infinite volume.

\begin{figure}
\vspace{-.2cm}
\centerline{\epsfxsize=0.40\textwidth\epsfbox{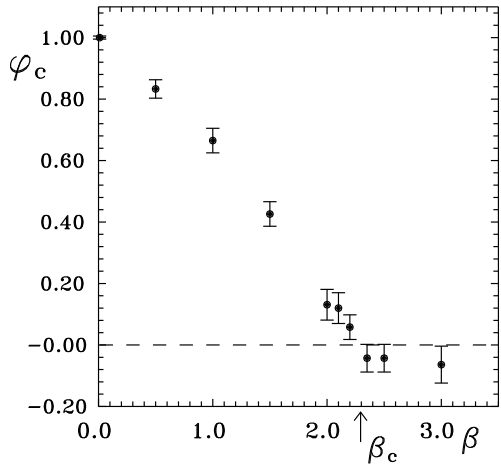}}
\vspace{-2.8cm}
\caption{The position of the minimum of the effective constraint
potential for the monopole creation operator.}
{~}
\vspace{-1.cm}
\end{figure}

The monopole creation operator \cite{KeKi86}
in the monopole current representation is studied in ref.\cite{Nak96}.
First the monopole action is reconstructed from the monopole currents in the
MA projection, and after that the expectation value of the monopole creation
operator is calculated in the quantum theory of monopole currents. Again,
the monopole creation operator depends on the temperature as the disorder
parameter.

\subsection{Effective Monopole Action}

The examples discussed in Sect. 3.1 show that there
exists the monopole condensate in the confinement phase of the lattice
gluodynamics. Thus the simplest (i.e., with the minimal number of
derivatives) effective Lagrangian for the abelian fields (diagonal gluon
fields) should be equivalent to the Lagrangian of the dual Abelian Higgs 
model. In this model, the confinement of quarks exists at the classical 
level. It is important to find out whether the effective Lagrangian for the 
abelian field (diagonal gluon field) in the continuum limit ($\beta \to 
\infty$) is close to the Lagrangian of the Abelian Higgs model. The effective 
Lagrangian for the monopole currents  
can be more easily reconstructed from the numerical data  \cite{Sul96}. For 
the $SU(2)$ lattice gauge theory in the MA projection the coefficients of 
the Lagrangian for the extended monopoles \cite{IvPoPo90} seem to 
scale \cite{Sul96}, which means that there exists a continuum limit of the 
effective action. Some preliminary results have been 
obtained by a similar study of the monopole Lagrangian for the lattice 
$SU(3)$ gluodynamics \cite{Sul96}.

\section{MONOPOLES AND INSTANTONS}

Since monopoles are responsible for the confinement, it is
important to find a general class of nonabelian fields which
generate monopoles, in particular, in the MA projection. This is a rather
complicated problem. There exists
the projection independent definition of the monopole
\cite{Faber}. Still it is unclear how these monopoles are related
to the abelian monopoles and what is the confinement mechanism if
monopoles are nonabelian. As claimed in \cite{GoMo96}, there
are some classical structures (bumps in the field
strength) which are correlated with the string tension. It is unclear
wether these structures are related to monopoles.

At present, there exists only one carefully studied example:
it occurs that instantons and monopoles in the MA projection are
interrelated. The simplest solution \cite{ChGu95} of the problem, shown
in Fig.~4, consists of the straight line monopole trajectory which goes
through the center of the instanton.

\begin{figure}[htb]
\vspace{-.3cm}
\centerline{\epsfxsize=0.32\textwidth\epsfbox{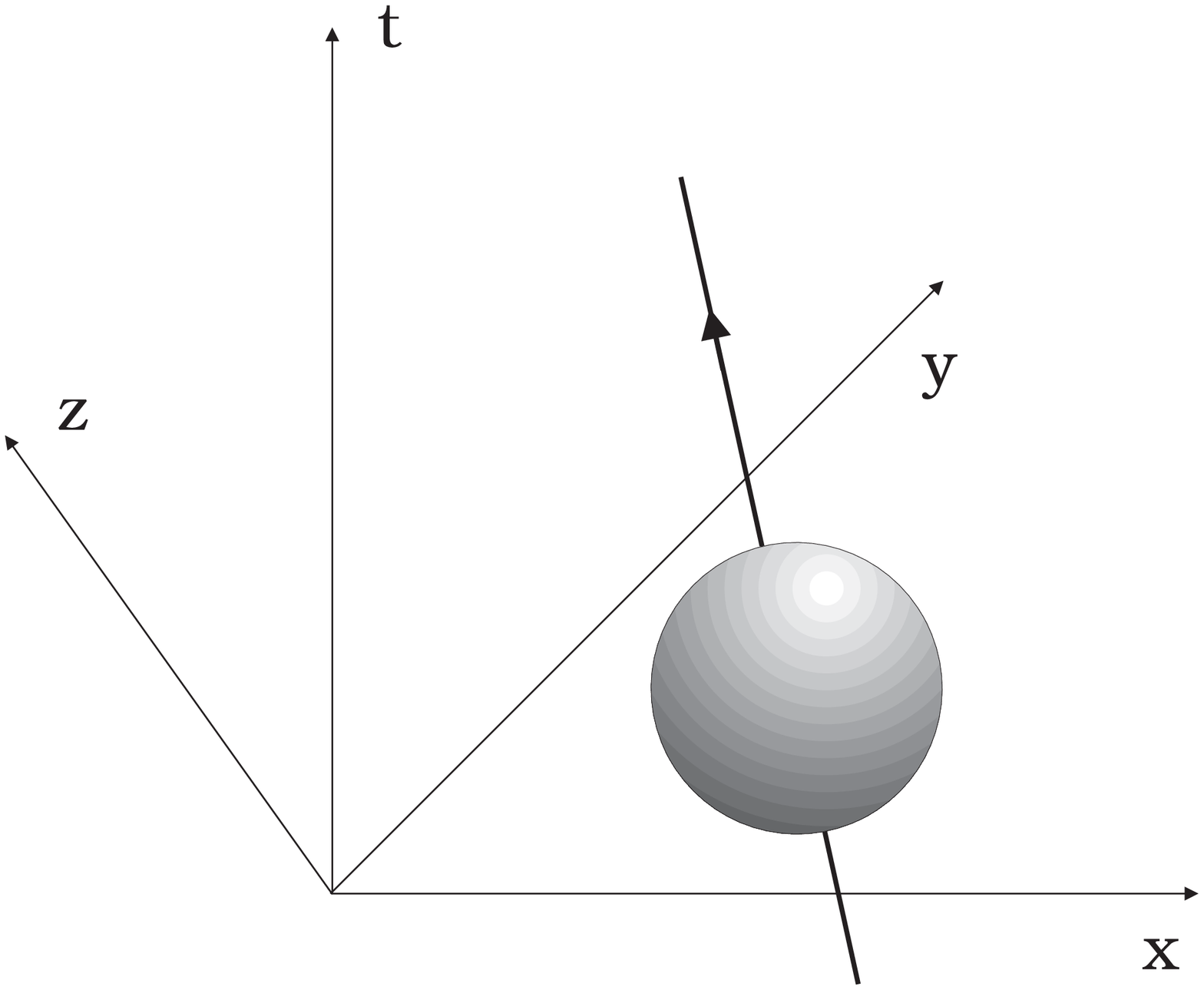}}
\vspace{-0.6cm}
\caption{Mononopole current in the instanton field. Sphere represents an
instanton, line is the monopole current.}

\vspace{-0.6cm}
\end{figure}
\newpage
A more complicated solution \cite{BrOrTa96}, shown in Fig.~5,
consists of the circular monopole current of radius $R$, and the
instanton of the width $\rho$ at the center of the monopole trajectory.

\begin{figure}[htb]
\centerline{\epsfxsize=0.32\textwidth\epsfbox{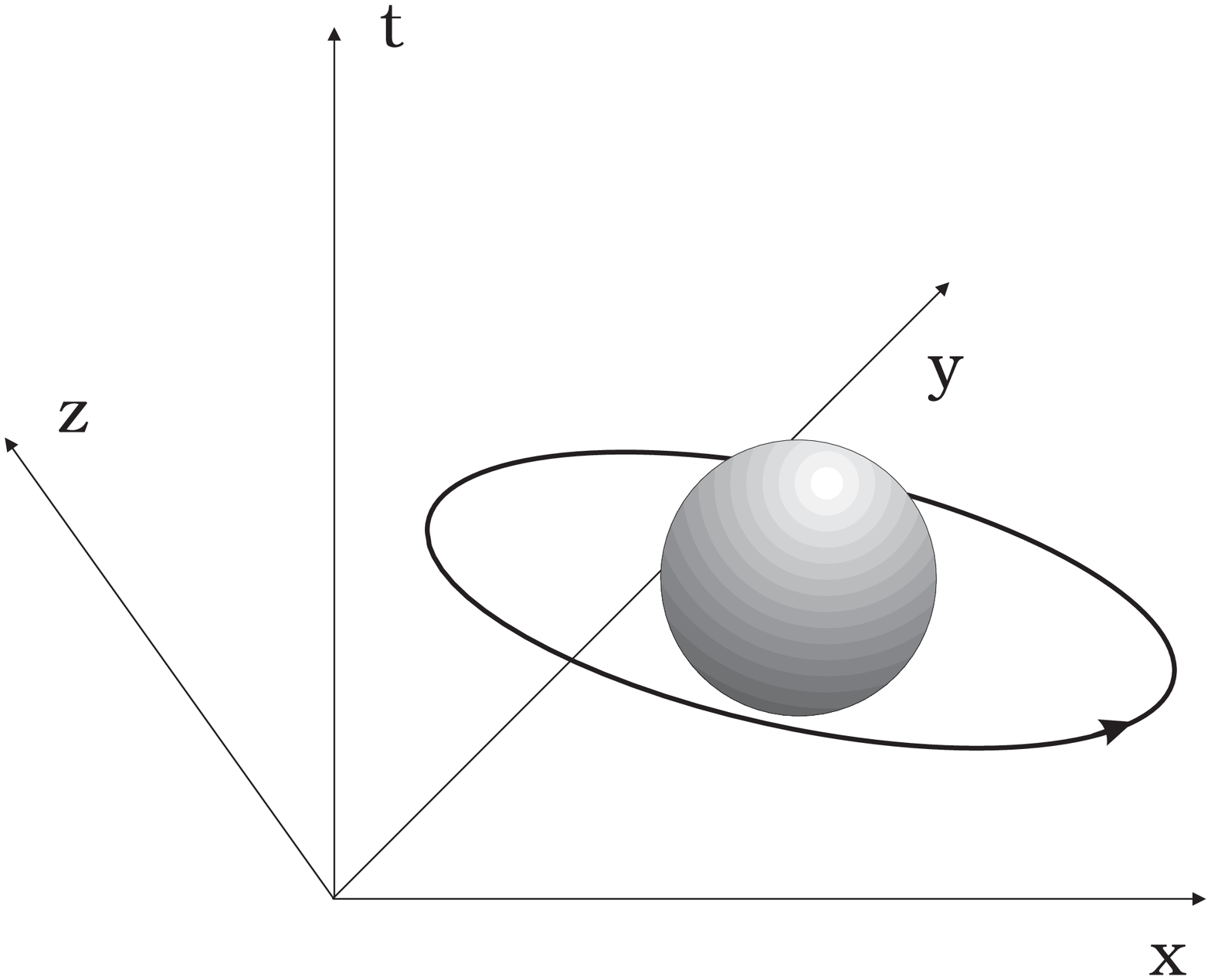}}
\vspace{-0.6cm}
\caption{Another solution of eq. (2),
the notations are the same as in Fig.~4.}
\vspace{ -0.8cm}
\end{figure}

It occurs that both solutions satisfy equation \eq{dR}, but the
minimization condition of \eq{R} holds only for the second
solution (Fig.~5) in the limit $R \to 0$.  It is important that the
leading correction to the minimization condition of \eq{R} is of the
order of $(R/\rho)^4\ln(R/\rho)$ \cite{BrOrTa96}. Therefore any small
quantum fluctuations (or even the coarseness of the lattice) can
create small monopole loops.

The solutions, shown in Figs.~4,5 are observed for the $SU(2)$
instantons in the MA projection on the lattice
\cite{HaTe95,GoPo95,BoSc96}. Moreover, there exists a correlation
between the topological charge and the monopole currents for the
cooled and non--cooled lattice field configurations in the MA
projection \cite{SuTaSaMi95,FeMaTh96}.  The relation observed between
instantons and monopoles allows one to discuss instanton--monopole models
of the QCD vacuum \cite{FuSaSu96}.

In the MA projection the nonabelian part of the gluon field is
suppressed and, therefore, the abelian field corresponding to the 
instanton field is almost selfdual and the magnetic current $j_\mu = \frac 
12 \partial_\nu \varepsilon_{\mu\nu\alpha\beta} f_{\alpha\beta}$ should be 
accompanied by the electric current $j_\mu =
\partial_\nu f_{\mu\nu}$.
This effect -- the dyon creation
by the instanton field -- is observed for the instantons on the
lattice \cite{BoSc96}. If dyons (not monopoles) are condensed, then
the Abrikosov--Nielsen--Olesen string dynamics may be very
nontrivial and the QCD strings may be fermionic (E.~Akhmedov, M.~Chernodub
and M.~Polikarpov, in preparation).

\section{AHARONOV -- BOHM EFFECT ON THE LATTICE}

There is a field theoretical analogue \cite{ABfield} of the
Aharonov-Bohm effect. The simplest example is the Abelian Higgs
theory. It is possible to represent the partition function of this
theory as a sum over closed surfaces \cite{ABAH}, which are the
world sheets of the Nielsen--Olesen strings: $\cZ = \sum_{\sigma}
\exp\{ - S(\sigma)\}$. In this representation the expectation value of
the Wilson loop for the charge $M$ is:

\beqn
<W_M(\cC)> = \frac{1}{\cZ} \sum_{\sigma}
\exp \{ - S_{local}(\sigma,\cC) \nonumber\\  
+ 2\pi i\frac MN \LL(\sigma,\cC) \}
\eeqn
Here $N$ is the charge of the Higgs field. The long--range interaction
described by the term which is proportional to the linking number
$\LL$ of the string world sheet $\sigma$ and the world line $\cC$
\footnote{In three dimensions there is the linking of
closed curves, the simplest example is shown in Fig.~6. In four
dimensions there exists the linking of a closed surface and a closed
curve, see Fig.~7.}  of the test charge is a four--dimensional
analogue of the Aharonov--Bohm effect: strings correspond to solenoids
which scatter charged particles.

\begin{figure}[htb]
\vspace{0.5cm}
\centerline{\epsfxsize=0.35\textwidth\epsfbox{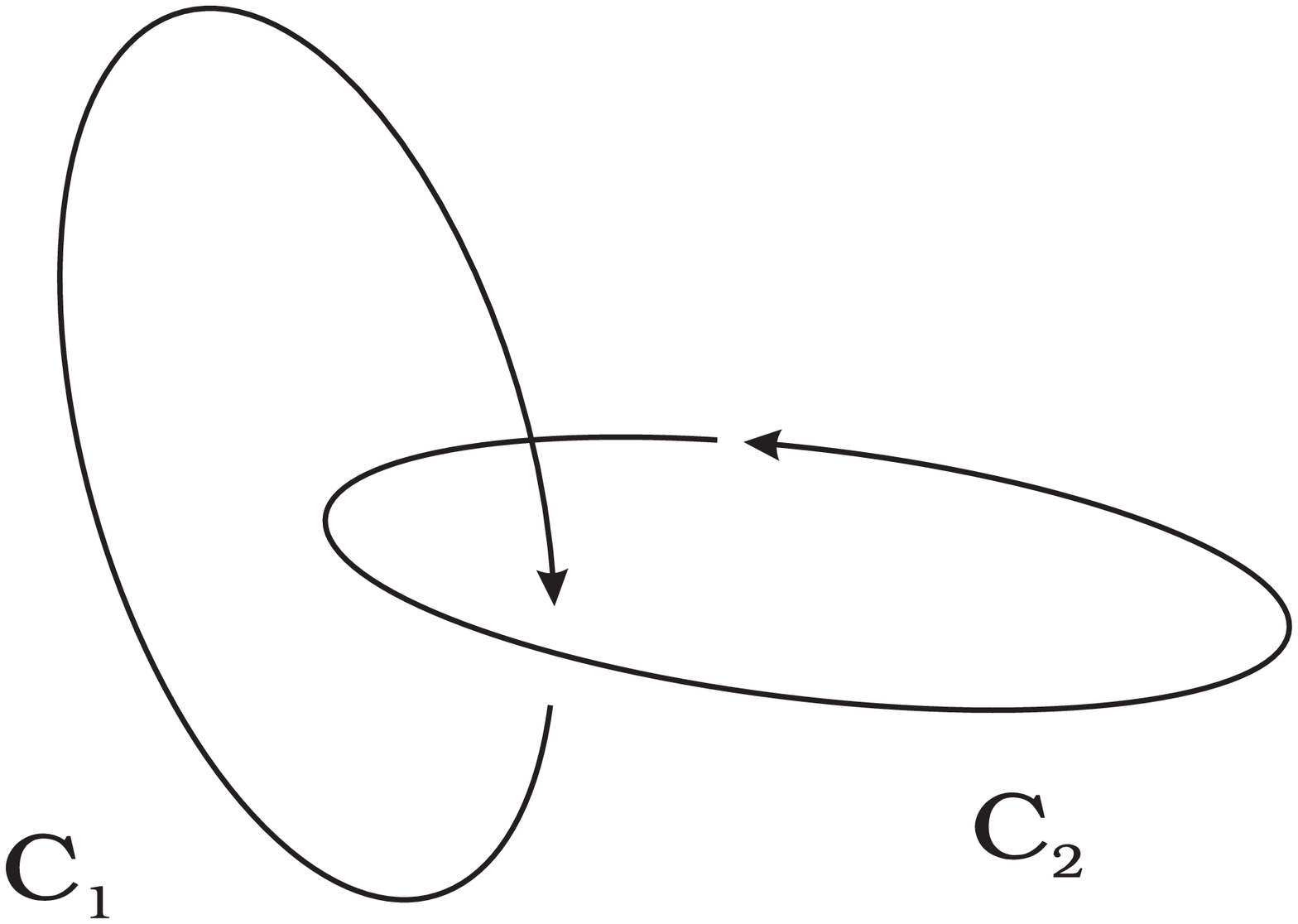}}
\caption{The linking of two curves $\cC_1$ and $\cC_2$
in three dimensions.}
~
\vspace{-0.8cm}
\end{figure}

\begin{figure}[htb]
\centerline{\epsfxsize=0.35\textwidth\epsfbox{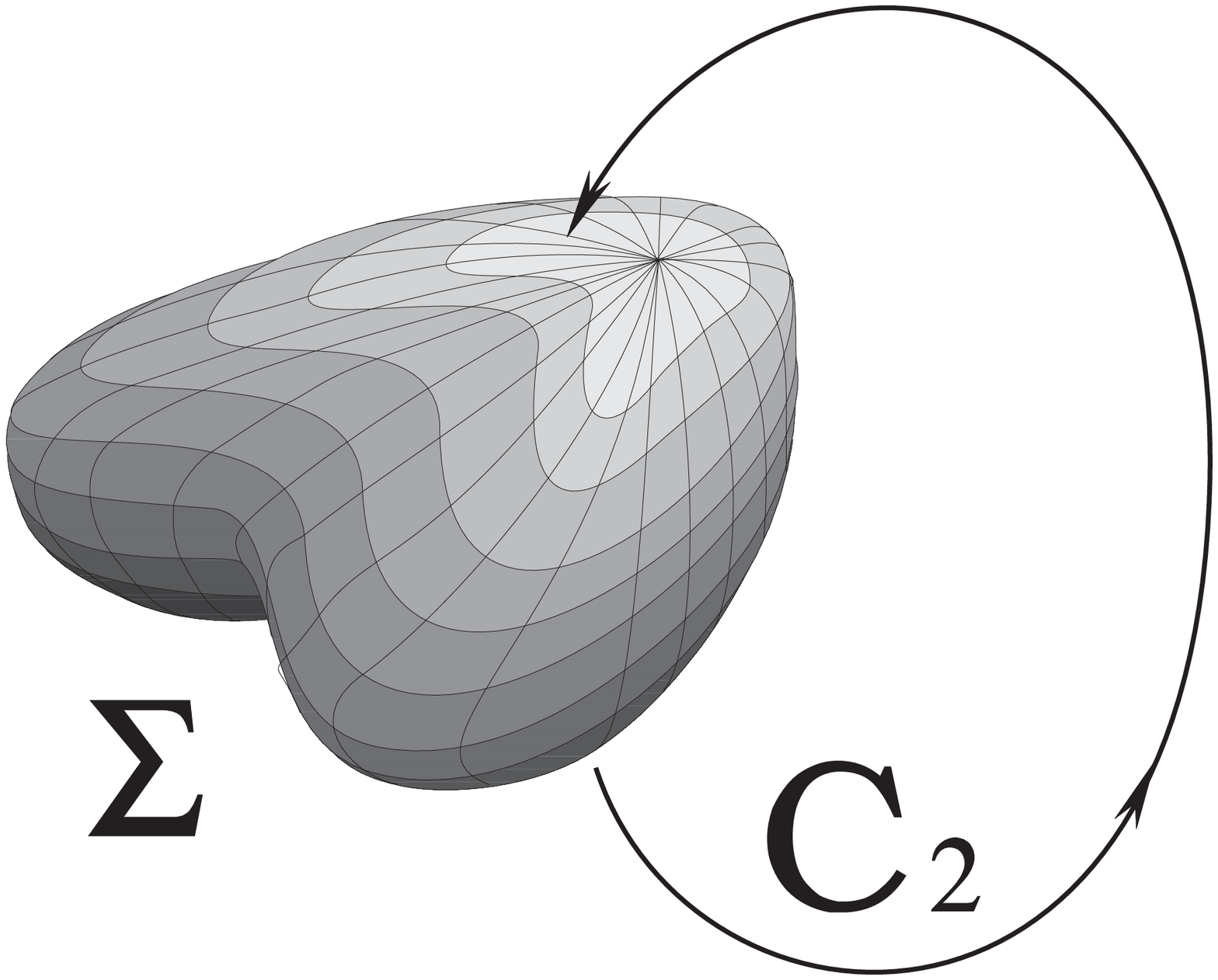}}
\caption{The linking of the curve $\cC_2$ and the closed surface $\Sigma$
in four dimensions.}
~
\vspace{-0.9cm}
\end{figure}

This topological interaction was found numerically \cite{ABn} in the
$3D$ Abelian Higgs model. In the abelian projection of gluodynamics
the off--diagonal gluons carry charge 2, the test quark in the
fundamental representation has charge 1, and the situation is quite
similar to the Abelian Higgs model. The problem is how to construct
the scalar field (an analogue of the Higgs field) from the vector
charged field (off--diagonal gluon). This can be done in several
ways. There are indications that for a new type of the abelian
projection of the $SU(2)$ lattice gluodynamics there exists
topological interaction (M.N.~Chernodub and M.I.~Polikarpov, work in
progress).  It occurs that

\beq
<AB> - <A>\cdot<B> \neq 0,
\eeq
where $A = W(\cC)$ is the Wilson loop, and $B = \exp\{\pi i
\LL(\sigma,\cC)\}$. The details will be given in a separate publication.

\section{CONCLUSIONS AND ACKNOWLEDGMENTS}

The facts described in this talk show that there are many
physical effects in the abelian projection of gluodynamics. It occurs that
in the MA projection (almost) all physical information is shifted to the
abelian part of the gluon field. Probably this is a nontrivial fact, which
is due to some small dynamical parameter.

I am very obliged to many people who supplied me with data
and helped me to prepare this small review.
My special thanks are to
V.~Bornyakov, R.~Brower, A.~Di~Giacomo, R.~Haymaker, H.~Markum,
G.~Poulis, A.~van~der~Sijs and T.~Suzuki. The help of my collaborators
M.~Chernodub, F.~Gubarev, A.~Veselov and E.~Akhmedov was very
important. This work is supported by the JSPS Program for Japan -- FSU
scientists collaboration, by the Grant INTAS-94-0840, and by the Grant
No. 96-02-17230a, financed by the Russian Foundation for Fundamental
Sciences.


\begin{thebibliography}{90}

\bibitem{tHo81}  G.~'t~Hooft, \NP, B190 [FS3] (1981) 455.

\bibitem{KrScWi87} A.S.~Kronfeld, M.L.~Laursen, G.~Schierholz and
U.J.~Wiese, Phys.Lett.198B
(1987) 516;\\
A.S.~Kronfeld, G.~Schierholz and U.J. Wiese, Nucl.Phys.
B293 (1987) 461.

\bibitem{MatH76} S.~Mandelstam, \PRp, 23C (1976) 245;\\
G.~{'t~Hooft}, "High Energy Physics", {\rm {Z}ichichi, Editrice
Compositori, Bolognia}, 1976.

\bibitem{SuYo90} T.~Suzuki and I.~Yotsuyanagi, \PR, D42 (1990) 4257.

\bibitem{HiKiKi91}  S.~Hioki et. al.,Phys.Lett., B272 (1991) 326;
ERRATUM-ibid. B281 (1992) 416.

\bibitem{StNeWe94} J.D.~Stack, S.D.~Neiman and R.J.~Wensley, Phys.Rev.,
D50 (1994) 3399.

\bibitem{BaBoMu96} G.S.~Bali, V.~Bornyakov, M.~Mueller-Preussker,
K.~Schilling; hep-lat/9603012, to be published in Phys.Rev.D.

\bibitem{StNe96}  J.D.~Stack, S.D.~Neiman, Phys.Lett., B377 (1996) 113.

\bibitem{StWeNe96} J.D.~Stack, R.J.~Wensley and S.D.~Neiman, preprint
OUTP-96-26-P, May 1996; hep-lat/9605045;\\
J.D.~Stack, these proceedings; hep-lat/9607064.

\bibitem{Yee94} K.~Yee, Nucl. Phys. B (Proc. Suppl.), 34 (1994) 189.

\bibitem{Kit96} S.~Kitahara, preprint KANAZAWA-96-12,
these proceedings.

\bibitem{Eji96} S.~Ejiri, these proceedings; hep-lat/9608001.

\bibitem{Suzm96} T.~Suzuki, S.~Kitahara, T.~Okude, F.~Shoji,
K.~Moroda and O.~Miyamura, Nucl. Phys. B (Proc. Suppl.) 47 (1996) 374,
hep-lat/9509016;\\ 
T.~Suzuki, preprint KANAZAWA-96-08, Jun 1996,
hep-lat/9606015.

\bibitem{DeFaGr96} L.~Del~Debbio, M.~Faber, J.~Greensite and S.~Olejnik,
Phys.Rev., D53 (1996) 5891.

\bibitem{Gre96}  L.~Del~Debbio, M.~Faber, J.~Greensite and S.~Olejnik,
these proceedings; hep-lat/9607053.

\bibitem{Pou96} G.~Poulis, preprint NIKHEF-95-064, Jan 1996.;
hep-lat/9601013.

\bibitem{Poul96} G.~Poulis, these proceedings, hep-lat/9608054.

\bibitem{ChVePo95} M.N.~Chernodub, M.I.~Polikarpov and A.I.~Veselov,
Phys. Lett. B342 (1995) 303.

\bibitem{IvPoPo90}  T.L.~Ivanenko, A.V.~Pochinskii and M.I.~Polikarpov,
Phys.Lett. B252 (1990) 631.

\bibitem{DiGi95} L.~Del Debbio, A.~Di~Giacomo, G.~Paffuti and P.~Pieri,
Phys.Lett. B355 (1995) 255.

\bibitem{HaTe96}  A.~Hart and M.~Teper, preprint OUTP-96-36-P;
hep-lat/9606022.

\bibitem{Hay96} K.~Bernstein, G.~Di~Cecio and R.W.~Haymaker, preprint
LSUHE-213-1996, Jun 1996, hep-lat/9606018; \\
K.~Bernstein, G.~Di~Cecio and R.W.~Haymaker, these proceedings;
hep-lat/9607075.

\bibitem{Sul96} T.~Suzuki et.al., these proceedings; hep-lat/9607054.

\bibitem{vdS96} A.J.~van~der~Sijs, these proceedings; hep-lat/9608041.

\bibitem{DiGir96} A.~Di~Giacomo, preprint IFUP-TH-31-96; hep-lat/9606001.

\bibitem{ChPoVeo95} M.N. Chernodub, M.I. Polikarpov and A.I. Veselov,
Nuclear Physics B (Proc. Suppl.) 49 (1996) 307; hep-lat/9512030.

\bibitem{Nak96} N.~Nakamura, V.~Bornyakov, S.~Ejiri,
S.~Kitahara, Y.~Matsubara and T.~Suzuki, these proceedings;
hep-lat/9608004.

\bibitem{CeGiPa96} G.~Di~Cecio, A.~Di~Giacomo, G.~Paffuti and M.~Trigiante,
preprint IFUP-TH-13-96, Mar 1996, cond-mat/9603139;\\
G.~Di Cecio, A.~Di Giacomo, G.~Paffuti and M. Trigiante, these
proceedings; hep-lat/9608014.

\bibitem{FrMa86} J.~Fr\"ohlich and P.A.~Marchetti,
Europhys. Lett. 2 ( 1986) 933.

\bibitem{PoPoWi91} M.I.~Polikarpov, L.~Polley and U.J.~Wiese,
Phys.Lett.B253 (1991) 212.

\bibitem{KeKi86} T.~Kennedy and C.~King, Comm.Math.Phys., 104 (1986)
327.

\bibitem{Faber}  M.~Zach, M.~Faber and P.~Skala, these proceedings;
hep-lat/9608009.

\bibitem{GoMo96}  A.~Gonzalez-Arroyo and A.~Montero, preprint FTUAM/96-13,
hep-th/9604017; \\
A.~Gonzalez-Arroyo and A.~Montero, these proceedings, hep-lat/9608035.

\bibitem{ChGu95}  M.N.~Chernodub and F.V.~Gubarev, JETP Lett.62 (1995)
100; hep-th/9506026.

\bibitem{BrOrTa96} R.C.~Brower, K.N.~Orginos and C.I.~Tan,
these proceedings; hep-lat/9608012.

\bibitem{HaTe95} A.~Hart and M.~Teper, preprint OUTP-95-44-P; hep-lat/9511016.

\bibitem{GoPo95} A.~Gonzalez-Arroyo and M.I.~Polikarpov, unpublished.

\bibitem{BoSc96} V.~Bornyakov and G.~Schierholz, preprint DESY 96-069;
hep-lat/9605019.

\bibitem{SuTaSaMi95} H.~Suganuma, A.~Tanaka, S.~Sasaki and
O.~Miyamura, in Proceedings of the LATTICE 95 symposium, Nucl.
Phys. B (Proc. Suppl.) 47 (1996); hep-lat/9512024.

\bibitem{FeMaTh96} M.~Feurstein, H.~Markum and St.~Thurner, these
proceedings, hep-lat/960803; \\
M.~Feurstein, H.~Markum and St.~Thurner, these proceedings,
hep-lat/9608038.

\bibitem{FuSaSu96} M.~Fukushima, S.~Sasaki, H.~Suganuma, A.~Tanaka, H.~Toki
and D.~Diakonov, hep-lat/9608084.

\bibitem{ABfield} M.G.~Alford and F.Wilczek, \PRL, 62 (1989) 1071;\\
M.G.~Alford, J.~March--Russel and F.Wilczek, \NP, B337 (1990) 695;\\
J.~Preskill and L.M.~Krauss, \NP, B341 (1990) 50;\\
M.~Alford, K-M.~Lee, J.~March-Russel and J.~Preskill, \NP,
B384 (1992) 251.

\bibitem{ABAH}  M.I.~Polikarpov, U.J.~Wiese and M.A.~Zubkov,
Phys.Lett. B309 (1993) 133. \\
E.T.~Akhmedov, M.N.~Chernodub, M.I.~Polikarpov and M.A.~Zubkov,
Phys.Rev. D53 (1996) 2087-2095.

\bibitem{ABn}  M.N.~Chernodub, F.V.~Gubarev and M.I.~Polikarpov, preprint
ITEP-TH-29-96, May 1996, hep-lat/9607045; \\
M.N.~Chernodub, F.V.~Gubarev and M.I.~Polikarpov, these
proceedings; hep-lat/9608075.

\end{thebibliography}
\end{document}